\documentclass[amsmath,amssymb,nofootinbib]{revtex4}
\usepackage[latin1]{inputenc}
\usepackage{graphicx}
\usepackage{dcolumn}
\usepackage{bm}
\usepackage{color}
\newcommand{\cab}{\bar{C}_{AB}}

\newcommand{\cai}{\bar{C}_{AI}}

\begin{document}
\title{Secure communication in the twin paradox.}
\author{Juan Carlos Garc\'ia-Escart\'in}
\email{juagar@tel.uva.es}
\author{Pedro Chamorro-Posada}
\affiliation{Dpto. de Teor\'ia de la Se\~{n}al y Comunicaciones. ETSI de Telecomunicaci\'on. Universidad de Valladolid. Campus Miguel Delibes. Paseo Bel\'en 15. 47011 Valladolid. Spain.}
\date{\today}
\begin{abstract}
The amount of information that can transmitted through a noisy channel is affected by relativistic effects. Under the presence of a fixed noise at the receiver, there appears an asymmetry between ``slowly aging'' and ``fast aging'' observers which can be used to have private communication transmission. We discuss some models for users inside gravitational wells and in the twin paradox scenario.  
\end{abstract}
\maketitle 
\vspace{-6ex}
\section{Communication and Relativity}
Noise limits the amount of information that can be sent through a communication channel. Shannon's noisy channel theorem establishes the maximum data rate for which reliable transmission is possible \cite{Sha49a}. This rate is the channel capacity, $C$. The capacity bound is supposed to give an absolute limit for communication which does not depend on a concrete transmission system.

However, the derivation of the traditional channel capacity does not take into account modern physical theories like Relativity and Quantum Mechanics. In particular, when relativistic effects are considered the channel capacity is modified. Most interestingly, in a communication between users who age at different rates there appears an asymmetry in channel capacity \cite{JC81}. We show some basic scenarios in which this asymmetry can be used to create a private channel between slowly aging users. Under the presence of a fixed local noise the ``young'' users can extract meaningful data from the channel while faster aging eavesdroppers lose the signal against the background noise. We give an example of secure channels based on the gravitational redshift and propose an alternative version of the twin paradox in which two travelling twins can communicate freely in the presence of different inertial observers. This reveals an intriguing connection between Relativity and Information Theory which can help to explore the limits of both disciplines.

We consider the family of signals of duration $T$, bandwidth $W$ and power $P$. While, strictly, no signal can be limited in both time and frequency, these signals can approximate any real signal with any desired accuracy \cite{Sle76}. We take a simplified communication system with one transmitter, one ideal lossless channel and a receiver which introduces a noise of power $N$. For this channel, Shannon's formula gives a capacity of $C=W\log_2\left(1+SNR\right)$ bits per second, where $SNR=P/N$ is the signal-to-noise ratio. If we try to send more bits per second (use a data rate $R>C$), the receiver has almost zero probability of translating the signal into data. For any lower rate $R<C$ there is always an encoding which allows for essentially error-free transmission. 

We will assume both for the non-relativistic and relativistic cases that there is an additive white gaussian noise source at the receiver with constant power spectral density $N_0$. In this case $SNR=\frac{P}{N_0W}$. 

If the receiver, including the noise source, is moving with respect to the transmitter, Relativity predicts different signal parameters at the transmitter and the receiver. All the transformations can be written in terms of the Doppler factor $\alpha=f'/f$  which gives the ratio of the frequency at the receiver and the transmitter frames. Table \ref{frames} gives the relevant conversion factors. The primed variables correspond to the parameters as measured in the receiver. We assume that the receiver knows in advance the frequency shift and that it can collect all the signal so that any Doppler beaming is not relevant. 

\begin{table}
 \begin{tabular}{| c | c c c |}
\hline
\bf{Parameter}&\bf{Transmitter}&\phantom{aaaaaaaa}&\bf{Receiver}\\
\hline
Power&$P$&$\rightarrow$&$P'=\alpha^2 P$\phantom{\Huge{|}}\\
\hline
Bandwidth&$W$&$\rightarrow$&$W'=\alpha W$\phantom{\Huge{|}}\\
\hline
Rate&$R$&$\rightarrow$&$R'=\alpha R$\phantom{\Huge{|}}\\
\hline
Time&$T$&$\rightarrow$&$T'=\frac{T}{\alpha\phantom{\Huge{|}}}$\phantom{\Huge{|}}\\
\hline
\end{tabular}
\label{frames}
\caption{Communication channels with moving observers: The signal parameters show different values in the frames of a transmitter and a receiver in relative motion. The signals can be related using the Doppler factor $\alpha$, which is the ratio of the frequencies in both frames.}
\end{table}

This relativistic channel has a capacity $C=W\log_2\left(1+\alpha SNR \right)$ in the transmitter's frame \cite{JC81}. The change in the capacity formula comes fundamentally from the difference in the signal-to-noise ratio. The given relativistic channel capacity can be applied in a wide range of situations where $P'=\alpha^2P$. 

For a moving receiver with a constant velocity, the $\alpha^2$ factor of the signal power can be deduced by applying the Lorentz transformation to the electric and magnetic fields and observing the Poynting vector at the receiver \cite{Ein05,vBla84,LL75}. There is also an intuitive argument. Imagine our signal is made up of a certain number of photons $n_p$ of frequency $f$ which are generated in a time interval $\Delta T$. The energy of each photon is $E=hf$. At the receiver, the frequency of each photon is multiplied by a Doppler factor $\alpha$ and the total measured energy is $\alpha$ times greater. This happens during a time interval $\Delta T/\alpha$. The number of photons does not change. The net effect is a factor of $\alpha^2$ in the power. If $\alpha>1$, we have more energetic photons arriving at a faster rate. If $\alpha<1$, the receiver sees less energetic photons during a longer time and perceives a signal of smaller power. 

The results can be extended to arbitrary signals in the presence of gravitational Doppler factors. Consider two observers at rest at different points of a stationary gravitational field. In that case, $Edt=E'dt'$ is a constant of motion \cite{TW01,OST00,HSS06,RAO08}. The energy ratio for different observers is equal to the Doppler factor and is the inverse of the ratio between time intervals as measured in each frame $\left(\alpha=\frac{E'}{E}=\frac{dt}{dt'}\right)$. These are the same ratios we had in the photon example. As this conservation of energy is valid for light and matter, the relativistic channel capacity formula can be extended to signals of any nature. It will hold the same for radio or sound waves or even for cannonballs. 

Strictly speaking, the asymptotic results of the noisy channel theorem are only valid for signals where $T\to \infty$ and $\alpha$ is constant. For more complicated situations where the gravitational field is not stationary or the receiver is moving with a non-uniform velocity, we can calculate the average channel capacity assuming intervals of constant $\alpha$ for which we can use the relativistic Shannon formula. This average capacity can help us to derive bounds on the maximum number of bits a user can decode during a time $T$.
 
We study some of the simplest cases with two users Alice, $A$, and Bob, $B$, which have identical transmitters and receivers. We have one channel in each direction. We call the capacity of the channel from Alice to Bob $C_{AB}$ and the capacity from Bob to Alice $C_{BA}$.

In the simplest scenario with uniformly moving observers, the channel capacity is symmetric $(C_{AB}=C_{BA})$. This is immediate from the principle of Relativity. Asymmetries would reveal a preferred frame. For approaching observers $(\alpha>1)$ the channel capacity can be greater than the usual Shannon limit. However, receding observers find it more difficult to communicate.  

When there is an asymmetry in aging (each communication party measures a different proper time between common events) there is also an asymmetry in channel capacity.

\section{The redshift channel}
We can use this asymmetry to establish a private channel. A simple case is gravitational redshift security. If Alice is inside a gravitational potential well and Bob is outside, the signals from Alice to Bob suffer a redshift $(\alpha<1)$. Signals from Bob to Alice are blue-shifted $(\alpha>1)$. Here, Alice ages more slowly than Bob and $C_{BA}>C_{AB}$. 

\begin{figure}[ht!]
\centering
\includegraphics{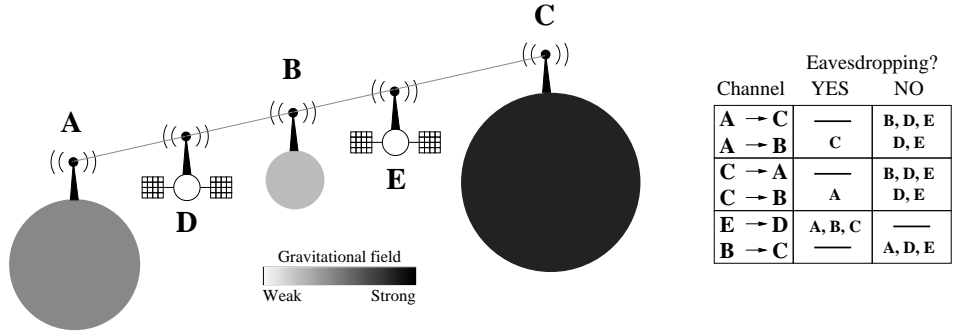}
\caption{\emph{Galactic Network:} If we take a series of stationary users on the surface of planets with different gravitational potentials, they will age at different rates. A user on any of the planets can always choose rates to communicate securely with slowly aging observers. If an ``older'' user inside a smaller gravitational well captures the signal, she cannot extract information from it. The table on the right shows some examples of the possible channels and the list of potential eavesdroppers. In the example, we suppose the local noise is the same in all the receivers and that an eavesdropper can always capture the whole transmitted signal. An eavesdropper near a black hole or inside any other strong gravitational field could intercept the signals and decode the data from most planets, even for a high local noise, but, as she approaches the event horizon, it becomes more and more difficult for her to send transmissions.}
\label{GalNet}
\end{figure}

Now imagine Alice and Bob are on the surface of two identical planets of equal mass and radius. In this case, there is no Doppler shift between Alice and Bob. However, a stationary eavesdropper, Eve, with the same receiver as Alice and Bob, but who is in deep space, far from any gravitational field, sees their signals shifted to red. Alice and Bob can establish a private channel by choosing a communication rate $R$ so that $C_{AB}=C_{BA}>R>C_{AE}=C_{BE}$. Signals that can be decoded by Alice and Bob look like noise to Eve. We can extend the model to a series of planets (See Figure \ref{GalNet}). 

\section{The twin channel}
By the equivalence principle, there will also be an asymmetry in the communication capacity of accelerating observers. It is not really necessary to have a uniform acceleration to achieve a capacity asymmetry. We present a simple example based on the twin paradox.
	
\begin{figure}[ht!]
\centering
\includegraphics{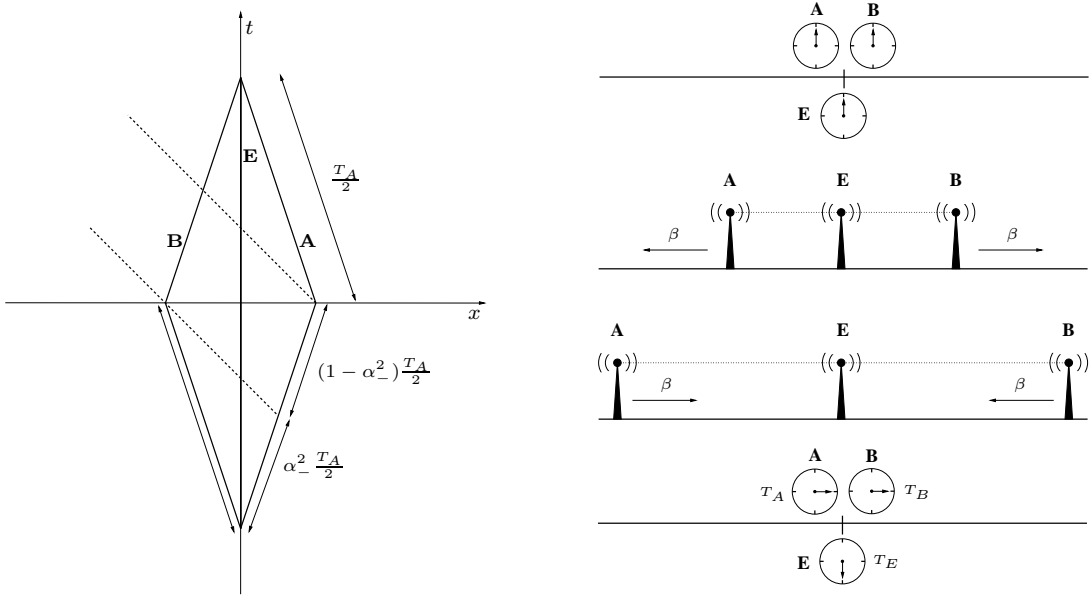}
\caption{\emph{Twin-paradox security:} Alice, Bob and  Eve start at the same point of space and synchronize their watches. Alice and Bob then move away in opossite directions with constant speed $\beta$ until they cover a certain distance and return to the origin at the same speed. When they all meet again, Alice's and Bob's watches are still synchronized, but Eve's watch shows a greater time $T_E>T_A=T_B$. During their journey, Alice and Bob can send and receive correctly up to $T_A\bar{C}_{AB}$ bits. Even if Eve intercepts all the transmission, she cannot decode more than $T_A\bar{C}_{AE}$ bits. This difference can be used to create a private channel between Alice and Bob.}
\label{TwinPar}
\end{figure}

We have triplets Alice, Bob and Eve who are originally in the same place. Alice and Bob move away from Eve at a constant velocity $\beta$ at opposite directions for a time $T_A/2$ and then return at the same velocity (see Figure \ref{TwinPar}). After the journey, the unaccelerated observer, Eve, will have aged more than Alice and Bob. The asymmetry appears as a result of the instantaneous acceleration at the turnover. 

The Doppler factors for the ``away'' and ``return'' journeys are $\alpha_{-}=\gamma\sqrt{1-\beta}$ and $\alpha_{+}=\gamma\sqrt{1+\beta}$.  The factor $\gamma=\frac{1}{\sqrt{1-\beta^2}}=\frac{T_E}{T_A}$ gives the ratio between the total time from the start of the journey to the reunion as measured by Eve ($T_E$) and as measured by Alice or Bob ($T_A$). The situation for Alice and Bob is totally symmetric with respect to Eve. With these factors we can calculate the average capacity of all the relevant channels (see appendices \ref{avcap} and \ref{security} for the details and complete proofs of security). 

The average channel capacity from Alice or Bob to Eve is:
\begin{equation}
\bar{C}_{AE}=\bar{C}_{BE}=\frac{W}{2}\left[log_2\left(1+\alpha_{-}SNR\right)+log_2\left(1+\alpha_{+}SNR\right)\right].
\label{cAE}
\end{equation}

The communication between Alice and Bob is divided in three stages with different Doppler factors. The average capacity of this channel is 
\begin{equation}
\bar{C}_{AB}=\bar{C}_{BA}=\frac{W}{2}\left[\alpha_{-}^2log_2\left(1+\alpha_{-}^2SNR\right)+\left(1-\alpha_{-}^2\right)log_2\left(1+SNR\right)+log_2\left(1+\alpha_{+}^2SNR\right)\right].
\label{cAB}
\end{equation}

It is easy to see that for $\gamma>1$, $\bar{C}_{AB}>\bar{C}_{BE}$. If there is an asymmetry in aging, there will always be a rate which allows private communication. 

During their journey, Alice and Bob will send each other more bits than Eve is able to decode correctly. This difference in private information can be later used to grow a secret key using privacy amplification \cite{BBC95}, like in quantum cryptography \cite{BB84}. Once Alice and Bob share a secret key, they can communicate securely using the perfect encryption method of the one-time pad \cite{Sha49b}.

There are also rates which give secure channels against families of eavesdroppers which have a Doppler factor of at most $\alpha_I$ with respect to the stationary Eve. We call an eavesdropper inertial if $\alpha_I$ is constant. Receivers moving in the $x$ axis with a constant speed with respect to Eve (and not crossing A, B or E) are inertial eavesdroppers. 

The average capacity of the channel with the inertial eavesdroppers is 
\begin{equation}
\bar{C}_{AI}=\frac{W}{2}\left[log_2\left(1+\alpha_{-}\alpha_I SNR\right)+log_2\left(1+\alpha_{+}\alpha_I SNR\right)\right].
\label{cAI}
\end{equation}
 
\begin{figure}[ht!]
\centering
\includegraphics{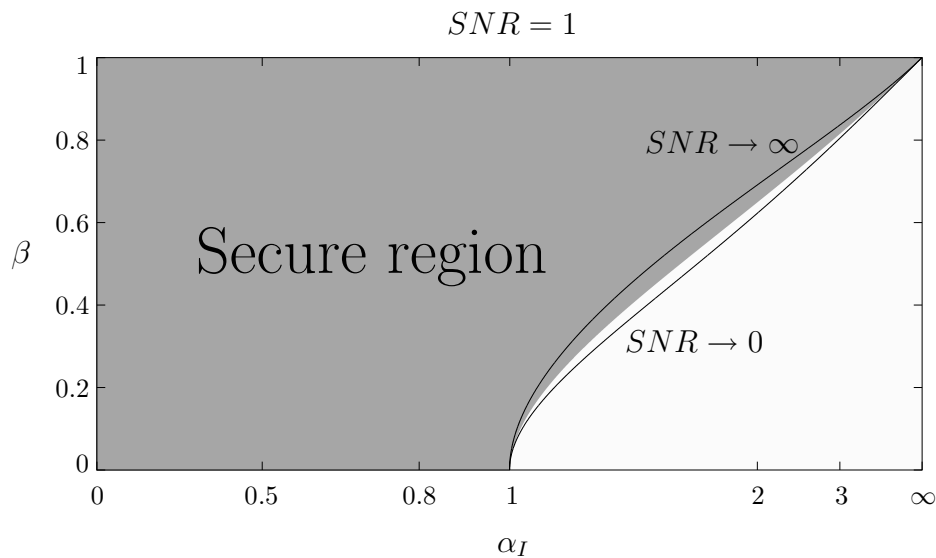}
\caption{\emph{Secure regions for the twin channel:} If Alice and Bob go away from a common point at speed $\beta$ and then come back at the same speed, their signals are secure against all the inertial eavesdroppers for which $\bar{C}_{AB}>\bar{C}_{AI}$. The shaded area corresponds to the secure region when the signal-to-noise ratio for $\beta=0$ is $SNR=1$. The curves show the border of the secure region for the cases of weak noise ($SNR\to\infty$) and strong noise ($SNR\to 0$). We have assumed that the inertial eavesdropper has the same local noise as the legitimate receiver. However, the graph gives a way to extend the scheme to different noise levels. A stationary eavesdropper for which (${N_0}'<N_0$) is equivalent to a moving observer with $\alpha_I=\frac{N_0}{{N_0}'}$. For any temperature greater than absolute zero there will be a local thermal noise with $N_0\neq 0$. If we can estimate a physical bound for the local noise in the eavesdropper's receiver, we can choose an appropriate speed $\beta$ which gives a secure channel. Unfortunately, even for small factors $\alpha_I$, $\beta$ must be close to the speed of light.}
\label{inert}
\end{figure}

Figure \ref{inert} shows the secure regions for different twin velocities $\beta$. For higher velocities  (and higher aging asymmetries), a smaller number of inertial observers are able to read the signals. These results can be generalized to arbitrary observers with a bounded Doppler factor and arbitrary noise levels in the eavesdropper (see Appendices \ref{arbdir} and \ref{arbnoise}).
	
The examples we have presented stress the curious asymmetries in communication that appear when there is differential aging. Choosing appropriate encodings and rates, slowly aging users can communicate efficiently while faster aging eavesdroppers can only extract a vanishing amount of data even if they capture the whole signal. As long as there is a minimum noise at the eavesdropper's receiver, there will be speeds or gravitational potentials which can be used to keep security.

Our relativistic cryptography scheme contrasts with the previous model based on the Unruh effect in which accelerating (slowly aging) eavesdroppers could not read the channel of fast aging user \cite{BHP09}. The difference comes from considering thermal noise at the receivers as the dominant source of noise. A complete relativistic channel description should include both cases. This opens an interesting line of research on the connection of Communications, Information and Relativity.

\section*{Acknowledgments}
Part of this work was done during the stay of Juan Carlos Garc\'ia Escart\'in at the Centre for Quantum Computation at Cambridge University (UK), which was funded by mobility program Jos\'e Castillejo Grant Ref. JC2009-00271. Juan Carlos would like to thank all the people there for their kind hospitality.

\section{APPENDICES}
\subsection{Average capacity}
In the twin paradox scenario, the channel changes with time. In this case, we can no longer use Shannon's noisy channel formula directly. We will use instead an average capacity $\bar{C}$. We define the average capacity in terms of the maximum number of bits that can be sent with our signal of length $T$ (for time and other parameters measured in the transmitter's frame).

The average capacity is chosen so that $N=T\bar{C}$, where $N$ is the maximum number of decodable bits that can be sent through the channel. We assume that the whole length of the transmission $T$ is divided into $n$ sections $T_i$ so that $T=\sum_{i=1}^{n}T_i$. During each segment of duration $T_i$ the channel is constant and we can assign it a capacity $C_i=W\log_2 \left(1+\alpha_i SNR \right)$. We suppose that each $T_i$ is long enough for the asymptotic results of Shannon's theorem to hold. 

In each of these segments the transmitter can find codes to send up to $N_i=T_iC_i$ bits. If Alice uses $n$ separate signals (one for each segment), she can send a total of $N=\sum_{i=1}^{n}N_i$ bits. This is not necessarily the best she can do. There might be, for instance, better encoding schemes using blocks of length $T$. However, for $n$ concatenated blocks of length $T_i$, for $i=1\ldots n$, she cannot send more information. 

With all these details in consideration, we define the average capacity as
\begin{equation}
\bar{C}=\frac{1}{T}\sum_{i=1}^{n}T_iC_i.
\label{avC}
\end{equation}
We can define the capacity of the channels going from Alice to different frames using the same formula. In this case, the $C_i$ would have a different Doppler factor $\alpha_i$ and would give a different maximum number of bits. We could even have a different number of segments in each frame. For different receiver frames the passage of time at the destination is also different. However, in the relativistic capacity formula of Jarett and Cover \cite{JC81} the relevant frame is that of the transmitter and only $\alpha_i$ changes with the receiver. When calculating the average capacity, we only need to calculate the time segments at the transmitter for which the signals will be transformed by a different Doppler factor when they arrive at the receiver. 

If Alice sends the same signal to two identical receivers, Bob and Eve, in different frames, the channel from Alice to Bob and from Alice to Eve will, in general, have different average capacities $\bar{C}_{AB}$ and $\bar{C}_{AE}$. There is an important difference between two channels with different capacities $C_{AB}>C_{AE}$ and two channels with different average capacities $\bar{C}_{AB}>\bar{C}_{AE}$. In the first case we can choose a rate $R$ so that $C_{AB}>R>C_{AE}$ to have a direct secure channel. At that rate, Bob can decode all the information and Eve can only get a vanishing amount of bits. 

In the second case, when $\bar{C}_{AB}>\bar{C}_{AE}$, we can only guarantee that Eve will get, at most, $\bar{C}_{AE}T$ bits. It could even be the case that in various of the $n$ sections $\bar{C}_{AB_{i}}<\bar{C}_{AE_i}$ allowing Eve to read all the information. Different eavesdroppers could read different segments. However, if the average capacity of the channel between Alice and Bob is greater that the capacity of the channels to any eavesdropper, Alice and Bob can share more bits than Eve is able to capture. As long as Alice and Bob are sure they have more common information than a potential eavesdropper, they can use privacy amplification schemes to distill a completely secret key \cite{BBC95}. This key, when used in conjunction with a one-time pad \cite{Sha49b} allows them to have a perfectly secure channel.

\subsection{Definitions and useful identities}

In the analysis of the average capacity and the proofs of security it will be useful to have some intermediate results which relate different parameters that appear in the twin paradox. In this section, we give some of these identities.

\subsubsection{Composition of Doppler factors}
For a transmitter $A$ and a receiver $B$, the Doppler factor $\alpha_{AB}=\frac{f_B}{f_A}$ gives the ratio between the frequencies that are measured in each frame. In our average capacity results it is useful to write the parameters in various frames in terms of the Doppler factors between two particular reference systems. 

In the twin paradox scenario we are working on, all the given Doppler factors are between frames in uniform motion in the $x$ direction. For two such systems $\alpha_{AB}=\alpha_{BA}$ (there are no preferred inertial frames). When we have more than two systems, we can compose the Doppler factors by simply multiplying them.

\begin{figure}[ht!]
\centering
\includegraphics[height=6.2cm]{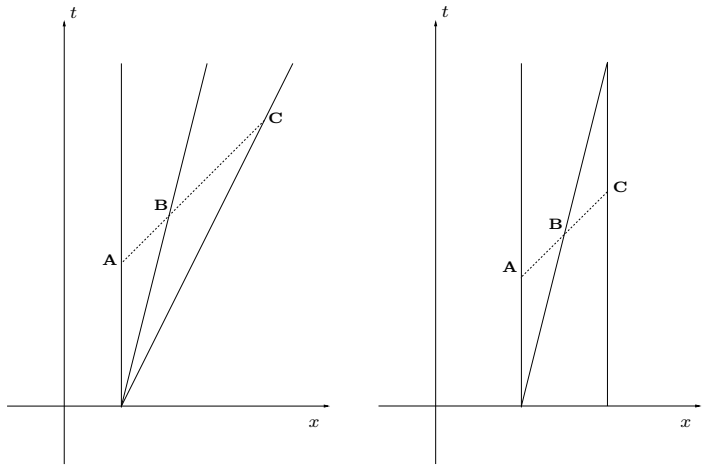}
\caption{Space-time diagrams for the Doppler factor identities.}
\label{doppcomp}
\end{figure}

We can see an example in Figure \ref{doppcomp} (left). We have three frames $A$, $B$ and $C$, and suppose $A$ is sending a signal at frequency $f_A$. Upon reception, $B$ sees a frequency $f_B=\alpha_{AB}f_A$ and $C$ sees $f_C=\alpha_{AC}f_A$. It is easy to see that

\begin{equation}
\alpha_{AC}=\frac{f_C}{f_A}=\frac{f_B}{f_A}\frac{f_C}{f_B}=\alpha_{AB}\alpha_{CB}.
\label{DopplerComp}
\end{equation}

There is one particular instance of this identity which will be useful in our calculations. We consider the Doppler factor of a system moving with a constant velocity $\beta$ either going away ($\alpha_{-}$) or approaching ($\alpha_{+}$) a reference frame we consider stationary. The space-time diagram of Figure \ref{doppcomp} (right) shows a scenario with two stationary observers, $A$ and $C$, and a travelling observer $B$ moving at velocity $\beta$ from $A$ to $C$. This is a particular case in which Equation (\ref{DopplerComp}) can be applied. The factor $\alpha_{AC}$ is clearly $1$ ($A$ and $C$ are the same reference frame) and $\alpha_{AB}=\alpha_{-}$ and $\alpha_{BC}=\alpha_{CB}=\alpha_{+}$. From Equation (\ref{DopplerComp}) we see that
\begin{equation}
\alpha_{+}\alpha_{-}=1.
\label{prodone}
\end{equation}

\subsubsection{Useful identities}
The Doppler factors $\alpha_{-}$ and $\alpha_{+}$ of the frames going away or approaching a stationary observer at a constant velocity $\beta$ are of a particular interest to us. They have different properties which will be useful when calculating the average capacities in the twin paradox communication scenario. 

We will also use the Lorentz factor $\gamma$ corresponding to a velocity $\beta$
\begin{equation}
\gamma=\frac{1}{\sqrt{1-\beta^2}}.
\end{equation}
Our velocities are normalized to the speed of light ($0\le |\beta| \le 1$) and, therefore, $\gamma \ge 1$. For $|\beta|\to 0$, $\gamma\to 1$ and for $|\beta|\to 1$, $\gamma \to \infty$. The function $\gamma(\beta)$ is increasing with $\beta$. If we have two velocities $\beta$ and $\beta'$ with Lorentz factors $\gamma$ and $\gamma'$ and $\beta>\beta'$, $\gamma>\gamma'$. The aging ratio between the stationary and the travelling observers in the twin paradox is the $\gamma$ corresponding to the velocity $\beta$ of the journey. 

There are different ways to write $\alpha_{-}$ and $\alpha_{+}$ in terms of $\beta$ and $\gamma$:
\begin{eqnarray}
\alpha_{+}=\sqrt{\frac{1+\beta}{1-\beta}},&&\alpha_{-}=\sqrt{\frac{1-\beta}{1+\beta}},\label{dopsqrt}\\
\alpha_{+}=\gamma(1+\beta),&&\alpha_{-}=\gamma(1-\beta).\label{dopgamma}
\end{eqnarray}

In the twin paradox, the description is given in terms of a positive velocity $\beta$ greater than zero (so that there is an aging asymmetry). For these $0<\beta \le 1$, we can see from (\ref{dopsqrt}) that
\begin{eqnarray}
\alpha_{+}> 1, &&\alpha_{-}< 1.\label{dopp1}\\
\end{eqnarray}
Factors $\alpha_{+}$ and $\alpha_{-}$ could only be equal if $\beta=0$ (when there is no aging difference).

Another relevant quantity is the sum $\alpha_{+}+\alpha_{-}$. From (\ref{dopgamma}), we have
\begin{equation}
\alpha_{+}+\alpha_{-}=\gamma(1+\beta)+\gamma(1-\beta)=2\gamma.
\label{doppsum}
\end{equation}
All the given expressions are valid for the Doppler factors $\alpha_{+}$ and $\alpha_{-}$ of any $\beta$.

\subsection{Average capacities in the twin paradox}
\label{avcap}
\subsubsection{Travelling twin to stationary eavesdropper}
\begin{figure}[ht!]
\centering
\includegraphics[height=6.2cm]{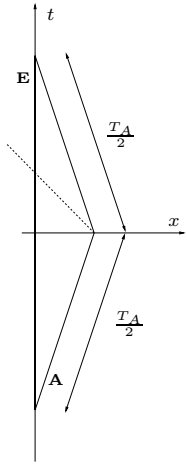}
\caption{Space-time diagram for the twin paradox channel going from a travelling Alice to a stay-at-home Eve.}
\label{stAE}
\end{figure}

First we will analyse the channel from a travelling Alice to a stay-at-home Eve. We have two equal segments of length $\frac{T_A}{2}$ and two Doppler factors, one for the away journey $\alpha_{-}=\sqrt{\frac{1-\beta}{1+\beta}}$ and one for the return journey $\alpha_{+}=\sqrt{\frac{1+\beta}{1-\beta}}$ (see Figure \ref{stAE}). 

From (\ref{avC}), we see that the average capacity is
\begin{equation}
\bar{C}_{AE}=\bar{C}_{BE}=\frac{W}{2}\left[log_2\left(1+\alpha_{-}SNR\right)+log_2\left(1+\alpha_{+}SNR\right)\right].
\label{cAE}
\end{equation}
This capacity is independent form the total journey time in Alice's frame, $T_A$. 

\subsubsection{Channel between travelling twins}
\begin{figure}[ht!]
\centering
\includegraphics{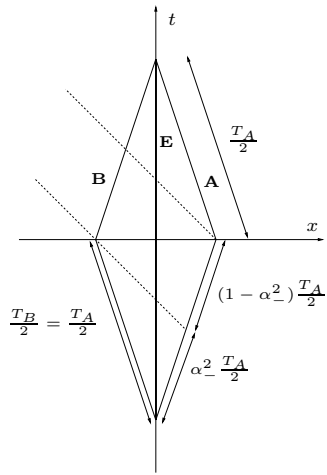}
\caption{Space-time diagram for two communicating travelling twins.}
\label{stAB}
\end{figure}

The second interesting channel is that connecting the travelling twins Alice and Bob. This channel is slightly more complicated. Now we have three time segments. In the first part of the communication, Alice and Bob are going away from Eve at velocity $\beta$. Their Doppler factor with respect to Eve is $\alpha_{-}$. They are also receeding from each other with a Doppler factor $\alpha_{-}'=\alpha_{-}^2$ between their frames. We can use Equation (\ref{DopplerComp}) and the reference factor with Eve to show that. Additionally, it is easy to see that $\alpha_{-}^2$ is the Doppler factor corresponding to two frames going away at speed $\beta'=\frac{2\beta}{1+\beta^2}$, which is the relativistic addition of the velocities of the frames of Alice and Bob. In the second part of the communication the frames are both moving to the right at velocity $\beta$ and there is no Doppler shift ($\alpha_{AB}=1$). Finally, in the third segment, both Alice and Bob approach Eve with velocity $\beta$. Using similar arguments to those of the outgoing journey, we can show that $\alpha_{AB}=\alpha_{+}^2$ in this final phase. Now we can compute the relevant $C_i$s. We only need the length of each segment, which can be obtained from the space-time diagram of the journey (Figure \ref{stAB}).

We will do all the calculations from Alice's frame, but everything is symmetric with respect to a transmitting Bob and a receiving Alice. The easiest segment is the third, which comprises the whole return journey of Alice ($T_3=\frac{T_A}{2}$). The first segment corresponds to the signals that arrive to Bob during the first half of his journey. In his frame, the interval lasts $\frac{T_B}{2}=\frac{T_A}{2}$ seconds\footnote{The times $T_A$ and $T_B$ must be equal. The journeys of Alice and Bob are equivalent. If they weren't, space wouldn't be isotropic.}. Time segments in different frames $a$ and $b$ are related by $T_b=\frac{T_a}{\alpha_{ab}}$. Therefore, $T_1=\alpha_{-}^2\frac{T_A}{2}$. The remaining time can be deduced by noticing that the two first communication segments take the first half of Alice's journey ($T_1+T_2=\frac{T_A}{2}$). 

Now we have $T_1=\alpha_{-}^2\frac{T_A}{2}$, $T_2=(1-\alpha_{-}^2)\frac{T_A}{2}$ and $T_3=\frac{T_A}{2}$, and the corresponding Doppler factors $\alpha_1=\alpha_{-}^2$, $\alpha_2=1$ and $\alpha_3=\alpha_{+}^2$ which define the capacity. The average capacity from Alice to Bob is 

\begin{equation}
\bar{C}_{AB}=\bar{C}_{BA}=\frac{W}{2}\left[\alpha_{-}^2log_2\left(1+\alpha_{-}^2SNR\right)+\left(1-\alpha_{-}^2\right)log_2\left(1+SNR\right)+log_2\left(1+\alpha_{+}^2SNR\right)\right].
\label{cAB}
\end{equation}
This capacity is also independent of $T_A$.

\subsubsection{Travelling twin to an inertial eavesdropper}
\begin{figure}[ht!]
\centering
\includegraphics{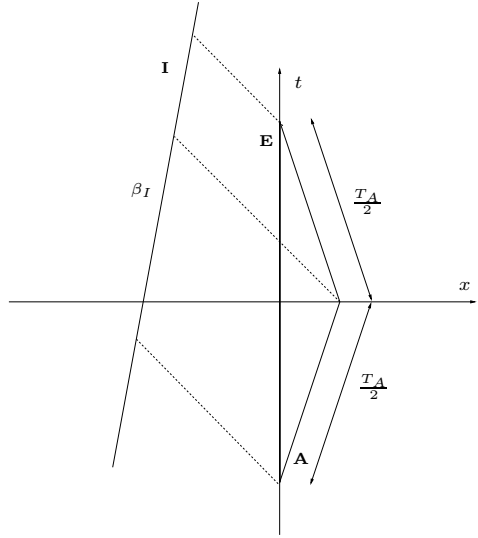}
\caption{Space-time diagram for the communication between Alice and an inertial eavesdropper.}
\label{stAI}
\end{figure}

A third relevant channel is the one between Alice and an inertial eavesdropper $I$ moving at a constant speed $\beta_{I}$ in the $x$ axis (Figure \ref{stAI}). We characterize the inertial observer by its Doppler factor $\alpha_{I}$ with respect to the stationary Eve. 

The analysis is exactly the same as in the case with a stationary observer. In Equation (\ref{cAE}) we only need to correct the Doppler factors multiplying them by $\alpha_{I}$. This is a direct consequence of the Doppler factor composition law of Equation (\ref{DopplerComp}). The average capacity is
\begin{equation}
\bar{C}_{AI}=\frac{W}{2}\left[log_2\left(1+\alpha_{-}\alpha_I SNR\right)+log_2\left(1+\alpha_{+}\alpha_I SNR\right)\right].
\label{cAI}
\end{equation}
Again, the capacity does not depend on $T_A$. For $\alpha_{I}=1$ (stationary observers), we recover capacity $\bar{C}_{AE}$. 

For certain gravitational fields, the Doppler factors coming from gravitation and motion can be factored \cite{RAO08}. In these cases, capacity $\bar{C}_{AI}$ is also valid for any observer in the $x$ axis with the same $\alpha_I$ as the inertial observer. The Doppler factor can come from a gravitational shift (either red or blue) or a combination of motion and gravitation.

\subsection{Proof of security}
\label{security}
\subsubsection{Security against a stationary eavesdropper}
The travelling twins can establish a private channel if they can share more information between them than with the frame of the eavesdropper. There is a secure channel if we can prove that $\bar{C}_{AB}>\bar{C}_{AE}$ (or that $\bar{C}_{AB}>\bar{C}_{AI}$ for inertial eavesdroppers). Equations (\ref{cAE}) and (\ref{cAB}) tell us that privacy is possible if,
\small
\begin{equation}
\frac{W}{2}\left[\alpha_{-}^2log_2\left(1+\alpha_{-}^2SNR\right)+\left(1-\alpha_{-}^2\right)log_2\left(1+SNR\right)+log_2\left(1+\alpha_{+}^2SNR\right)\right]>\frac{W}{2}\left[log_2\left(1+\alpha_{-}SNR\right)+log_2\left(1+\alpha_{+}SNR\right)\right].
\label{ineqBasic}
\end{equation}
\normalsize
We can simplify the expression extracting common factors and noticing that the logarithm is an increasing function. Proving that the channel between Alice and Bob is secure is equivalent to showing that
\begin{equation}
\alpha_{-}^2log_2\left(1+\alpha_{-}^2SNR\right)+\left(1-\alpha_{-}^2\right)log_2\left(1+SNR\right)+log_2\left(1+\alpha_{+}^2SNR\right)>log_2\left(1+\alpha_{-}SNR\right)+log_2\left(1+\alpha_{+}SNR\right),
\label{ineqLog}
\end{equation}
or
\begin{equation}
log_2\left(\left(1+\alpha_{-}^2SNR\right)^{\alpha_{-}^2}\left(1+SNR\right)^{1-\alpha_{-}^2}\left(1+\alpha_{+}^2SNR\right)\right)>log_2\left(\left(1+\alpha_{-}SNR\right)\left(1+\alpha_{+}SNR\right)\right),
\label{ineqProd}
\end{equation}
or
\begin{equation}
\left(1+\alpha_{-}^2SNR\right)^{\alpha_{-}^2}\left(1+SNR\right)^{1-\alpha_{-}^2}\left(1+\alpha_{+}^2SNR\right)>\left(1+\alpha_{-}SNR\right)\left(1+\alpha_{+}SNR\right).
\label{ineqArg}
\end{equation}

As $\alpha_{-}<1$, $1+SNR>1+\alpha_{-}^2SNR$. Using (\ref{prodone}), we can give a lower bound to the left hand side of (\ref{ineqArg}):
\begin{equation}
\left(1+\alpha_{-}^2SNR\right)^{\alpha_{-}^2}\left(1+SNR\right)^{1-\alpha_{-}^2}\left(1+\alpha_{+}^2SNR\right)\geq \left(1+\alpha_{-}^2SNR\right)\left(1+\alpha_{+}^2SNR\right)=1+\alpha_{+}^2SNR+\alpha_{-}^2SNR+SNR^2.
\end{equation}

Expanding also the right hand side of (\ref{ineqArg}), we can prove $\bar{C}_{AB}>\bar{C}_{AE}$ by showing 
\begin{equation}
1+\alpha_{+}^2SNR+\alpha_{-}^2SNR+SNR^2 > 1+ \alpha_{+}SNR+\alpha_{-}SNR+SNR^2.
\end{equation}
Taking out common factors, this reduces to proving 
\begin{equation}
\alpha_{+}^2+\alpha_{-}^2 >  \alpha_{+}+\alpha_{-},
\end{equation}
which, from (\ref{doppsum}), is equivalent to 
\begin{equation}
\label{gammaproof}
\gamma' > \gamma,
\end{equation}
for the $\gamma'$ corresponding to $\alpha_{+}'=\alpha_{+}^2$ and $\alpha_{-}'=\alpha_{-}^2$ (a velocity $\beta'=\frac{2\beta}{1+\beta^2})$. For $0<\beta<1$, $\beta'>\beta$. As $\gamma(\beta)$ is increasing, (\ref{gammaproof}) is true for all the $\beta$s we have in the twin paradox. $\blacksquare$  

\subsubsection{Security against inertial eavesdroppers}
\begin{figure}[ht!]
\centering
\includegraphics{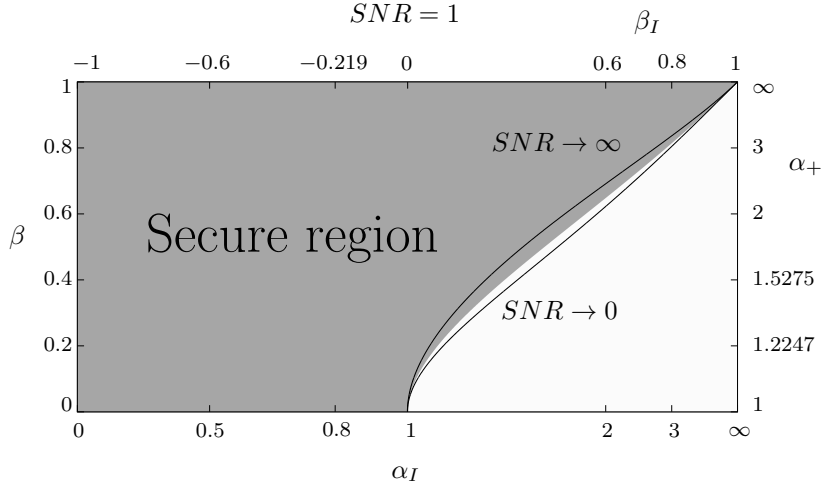}
\caption{\emph{Security regions for inertial eavesdroppers:} The limiting values, $SNR\to 0$ and $SNR\to \infty$, are derived in Section \ref{limit}.}
\label{SecureZones}
\end{figure}

The channel between Alice and Bob is also secure against a family of inertial eavesdroppers with an $\alpha_I$ below a certain threshold. If we take the expressions for $\bar{C}_{AB}$ and $\bar{C}_{AI}$, we can give a security condition similar to (\ref{ineqArg}). Alice and Bob can share a secure channel if
\begin{equation}
\left(1+\alpha_{-}^2SNR\right)^{\alpha_{-}^2}\left(1+SNR\right)^{1-\alpha_{-}^2}\left(1+\alpha_{+}^2SNR\right)>\left(1+\alpha_{I}\alpha_{-}SNR\right)\left(1+\alpha_{I}\alpha_{+}SNR\right).
\label{ineqArgI}
\end{equation}
For any given $SNR$ we can evaluate (\ref{ineqArgI}) numerically. Figure \ref{SecureZones}, which corresponds to Figure 3 in the paper, gives the secure regions (shaded) for $SNR=1$. In this Figure, we have also included the velocity values $\beta_I$ which give a Doppler factor $\alpha_I$ between the inertial observer and our reference stationary Eve. The factor $\alpha_{+}$ for the twin velocity $\beta$ is included for comparison. 

These maps of the secure communication regions against inertial eavesdroppers are a good starting point for many interesting generalizations. In the following sections, we will extend the security results to different signal-to-noise ratios and to eavesdroppers moving in any direction and with arbitrary noise levels. 

First, we will derive the analytic expressions for the frontiers of the secure regions for different signal-to-noise ratio regimes. Then, we will give a worst-case bound on the security against arbitrary inertial observers not moving in the $x$ direction. Finally, we will show how to extend the security scheme to situations where the eavesdropper has a different local noise level.

\subsection{Limiting cases}
\label{limit}
The secure regions give the speeds $\beta$ which guarantee the existence of a private channel for Alice and Bob for a given family of inertial observers with an $\alpha_I$ below a certain value. For $\alpha_I<1$ Alice and Bob always have a secure channel. The average capacity $\bar{C}_{AI}$ is bounded by $\bar{C}_{AE}$ which is always smaller than $\bar{C}_{AB}$. 

On the other hand, for a fixed twin velocity $\beta$ there is always an eavesdropper with $\alpha_{I}>1$ for which $\bar{C}_{AI}\ge\bar{C}_{AB}$. She can listen to the channel between Alice and Bob. Below that $\alpha_{I}$ the channel is secure. Above it, the eavesdropper might decode the transmission. The frontier depends on the value of the signal-to-noise ratio. Both $\cab$ and $\cai$ grow with $SNR$. However, for our range of interest with $\alpha_{I}>1$, smaller values of $SNR$ will reduce the amplifying effect of $\alpha_{I}$. 

An inertial eavesdropper with a Doppler factor $\alpha_{I}$ is equivalent to a stationary eavesdropper with signal-to-noise ratio $SNR'=\alpha_{I}SNR$. For the same increment in $\alpha_I$, a greater $SNR$ will give a greater increment in $SNR'$. This gives two important limiting cases. The first is the strong signal (low noise) limit, with $SNR\to \infty$. The second is the weak signal (high noise) limit with $SNR\to 0$. This last limit gives the maximum attainable secure region. Alice and Bob should try to work with small values of $SNR$ whenever possible.

\subsubsection{$SNR\to \infty$:}
We first study the worst case against inertial observers. In the strong signal limit we suppose that even for high velocities (very small $\alpha_{-}$) $\alpha_{-}^2 SNR\gg 1$. In that case, we can approximate the security condition (\ref{ineqArgI}) by
\begin{equation}
\left(\alpha_{-}^2SNR\right)^{\alpha_{-}^2}\left(SNR\right)^{1-\alpha_{-}^2}\left(\alpha_{+}^2SNR\right)>\left(\alpha_{I}\alpha_{-}SNR\right)\left(\alpha_{I}\alpha_{+}SNR\right),
\end{equation}
which is equivalent to asking that 
\begin{equation}
{\alpha_{-}^2}^{\alpha_{-}^2}\alpha_{+}^2>\alpha_{I}^2.
\end{equation}
If we use (\ref{prodone}), we can write everything in terms of $\alpha_{-}$ only. The condition for security in the strong signal limit becomes 
\begin{equation}
\alpha_{-}^{\alpha_{-}^2-1}>\alpha_{I}.
\end{equation}
This condition gives the limiting upper curve of Figure \ref{SecureZones}. When $\beta\to 1$ ($\alpha_{-}\to 0$) the condition can be approximated by
\begin{equation}
\alpha_{+}>\alpha_{I}.
\end{equation}
For high $\beta$s, an inertial eavesdropper must, at least, match the twins' velocity. For smaller values of $\beta$ a smaller $\beta_I$ is enough to compromise security.

\subsubsection{$SNR\to 0$:}
Similarly to the previous case, with this condition we mean that even for a high $\beta$ we have an $SNR$ so that $\alpha_{+}^2 SNR\ll 1$. In that case we can use the approximation $\displaystyle\lim_{x\to 0} \log_2(1+x)=\frac{1}{\ln 2}x$. We start from security condition (\ref{ineqLog}) adapted to the inertial observer case,
\begin{equation}
\alpha_{-}^2log_2\left(1+\alpha_{-}^2SNR\right)+\left(1-\alpha_{-}^2\right)log_2\left(1+SNR\right)+log_2\left(1+\alpha_{+}^2SNR\right)>log_2\left(1+\alpha_{I}\alpha_{-}SNR\right)+log_2\left(1+\alpha_{I}\alpha_{+}SNR\right).
\end{equation}
In the weak signal limit, the condition becomes 
\begin{equation}
\alpha_{-}^4SNR+\left(1-\alpha_{-}^2\right)SNR+\alpha_{+}^2SNR>\alpha_{I}\alpha_{-}SNR+\alpha_{I}\alpha_{+}SNR.
\label{strong}
\end{equation}
Taking out common factors we have the security condition
\begin{equation}
\frac{\alpha_{-}^4+1-\alpha_{-}^2+\alpha_{+}^2}{\alpha_{+}+\alpha_{-}}>\alpha_{I}.
\end{equation}
This formula could also be written only in terms of $\alpha_{-}$ or $\alpha_{+}$, but that doesn't provide any additional insight. Once again, in the $\beta\to 1$ limit ($\alpha_{-}\to 0$ and $\alpha_{+}\to \infty$) we recover the condition
\begin{equation}
\alpha_{+}>\alpha_{I}.
\label{condsecweak}
\end{equation}
When $\beta$ is close to the speed of light, the results for high and low $SNR$ converge. If $\alpha_I$ comes only from motion in the $x$ direction, $\beta>\beta_I$ guarantees security. 

\subsection{Inertial eavesdroppers in arbitrary directions}
\label{arbdir}
The security regions we have derived are valid for eavesdroppers in the $x$ axis with a Doppler factor $\alpha_{I}$ (caused either by a uniform movement or a gravitational field). In this section, we show how to generalize the security conditions to an inertial eavesdropper $I'$ moving in an arbitrary direction. If $I'$ has a velocity of absolute value of, at most, $|\beta_{I}'|=\beta_I$, her Doppler factor $\alpha_{I}'$ will be bounded by $\alpha_{+}\alpha_{I}$ (for the $\alpha_{+}$ and $\alpha_I$ of the previous sections). This bound limits the capacity of the eavesdropper's channels with Alice and Bob. We can adapt the security conditions so that $\bar{C}_{AB}$ is always greater than both $\bar{C}_{AI'}$ and $\bar{C}_{BI'}$.

To prove $\alpha_{I}'<\alpha_{+}\alpha_I$, we look into the worst case situation (highest $\alpha_{I}'$). The Doppler factor resulting from any uniform motion can be written as \cite{Ein05}
\begin{equation}
\alpha=\frac{1-\beta'\cos\Phi}{\sqrt{1-\beta'^2}}.
\label{generalDopp}
\end{equation}
Here, $\beta'$ is the relative velocity between the transmitter at the moment of transmission and the receiver at the moment of reception. $\Phi$ gives the angle between the direction of the velocity of the receiver and the direction of propagation of the signal, both measured at the moment of reception. This $\alpha$ is maximized for $\Phi=\pi$ (a receiver approaching an incoming wave).

We can study the relevant Doppler factors in this new situation from the relativistic composition of two velocities. We will discuss the communication with Alice, but everything will also hold for Bob. Suppose we know the velocities of the two observers in the frame of Eve. Observer $I'$ has a velocity $\beta_I$ in a direction given by the unit vector $\hat{u}_I$ and observer A has a velocity $\beta$ in the direction $\hat{u}$. For these conditions, the relative velocity between the observers in the resulting direction is \cite{Ein05}
\begin{equation}
\beta'=\frac{\sqrt{\beta_{I}^2+\beta^2+2\beta\beta_I\cos\Psi-\beta_I^2\beta^2\sin^{2}\Psi}}{1+\beta_I\beta\cos\Psi},
\label{generalAdd}
\end{equation}
where $\Psi$ is the angle between $\hat{u}_I$ and $\hat{u}$. We are interested in the maximum relative velocity between $A$ and $I'$. This velocity can help us to give a limit to the maximum attainable Doppler factor of a general inertial eavesdropper in the twin scenario. The dependence of $\Psi$ means that there are angles for which the same velocities give a higher relative velocity. We can easily find the optimal values. Taking the first derivative of $\beta'$ with respect to $\Psi$ we see that $\beta'$ becomes extremal if:

\begin{enumerate}
\item \emph{Either $\beta_I$ or $\beta$ are equal to $1$, for any value of $\Psi$ and the other velocity.} In this case, $\beta'=1$. This is a consequence of the constant value of $c$: a photon will be seen to move at the speed of light in any other inertial frame. 
\item \emph{Either $\beta_I$ or $\beta$ are equal to $0$, for any value of $\Psi$ and the other velocity.} Here, $\beta'$ is the other velocity. This case and the previous one are saddle points.
\item \emph{$\sin\Psi=0$, for any combination of velocities $\beta_I$ and $\beta$.} From the second derivative, we see that $\Psi=0$ gives a minimum and $\Psi=\pi$ a maximum. Velocity $\beta'$ is the one-dimensional relativistic addition of velocities. The maximum happens for two frames approaching in the same direction, as expected. This means that an inertial observer with velocity $\beta_I'\le \beta_I$ has a Doppler factor $\alpha_I'$ smaller than $\alpha_{+}\alpha_{I}$ (the factor for the addition of velocities $\beta$ and $\beta_I$ in the same direction). 
\end{enumerate}

The limiting case is an inertial observer approaching Eve in the $x$ direction. For such an eavesdropper, both Equation (\ref{generalDopp}) and Equation (\ref{generalAdd}) are maximized. For our $\alpha_{I'}<\alpha_{+}\alpha_{I}$, both $\bar{C}_{AI'}$ and $\bar{C}_{BI'}$ are bounded by $W\log_2\left(1+\alpha_{+}\alpha_{I}SNR\right)$. We can now adapt condition (\ref{ineqBasic}) to a general observer and write it as
 \begin{equation}
\frac{W}{2}\left[\alpha_{-}^2log_2\left(1+\alpha_{-}^2SNR\right)+\left(1-\alpha_{-}^2\right)log_2\left(1+SNR\right)+log_2\left(1+\alpha_{+}^2SNR\right)\right]>Wlog_2\left(1+\alpha_{+}\alpha_{I}SNR\right).
\label{ineqGendir}
\end{equation}

We can see this condition in the limit of weak signal, $SNR\to 0$, which was the regime which offered a greater security for the twin channel. The twins choose their transmitting power and can always place themselves in this situation. Under this approximation, we can derive a condition similar to inequality (\ref{strong}), which in our new scenario becomes 
\begin{equation}
\alpha_{-}^4SNR+\left(1-\alpha_{-}^2\right)SNR+\alpha_{+}^2SNR>2\alpha_{I}\alpha_{+}SNR.
\end{equation}
We can group all the terms relative to the twins' movement in the left to obtain
\begin{equation}
\alpha_{-}^5+\alpha_{-}-\alpha_{-}^3+\alpha_{+}>2\alpha_{I}.
\end{equation}
As $\alpha_{-}<1$, $\alpha_{-}-\alpha_{-}^3>0$. We can guarantee security if
\begin{equation}
\alpha_{+}>2\alpha_{I}.
\end{equation}
This bound is not optimal. The channel capacity will be different for the outgoing and return journeys. We can use that to our advantage to give a better bound. In a general case, we need to pay attention to the combination of velocities, directions and initial position of the eavesdroppers in order to compute the real capacity. This simpler crude bound suffices for our purposes. We can also include the effects of gravitation. For any particular case, we can find a maximum $\alpha_I'=\alpha_{+}\alpha_I$ and give the corresponding security condition.

\subsection{Eavesdroppers with an arbitrary local noise}
\label{arbnoise}
In our security conditions we have assumed the eavesdroppers have the same local noise as the legitimate users of the channel. In this section, we show that, as long as the eavesdropper has a receiver with a local noise greater than zero, there are velocities $\beta$ which allow Alice and Bob to preserve privacy. 

Imagine Alice and Bob have receivers with a local white noise of power spectral density $N_0$ and a stationary eavesdropper with a white noise of power spectral density $N_0'$ so that $0<N_0'<N_0$. The assumption of a non-zero gaussian noise is reasonable, as any receiver with a temperature greater than absolute zero will suffer some thermal noise. This stationary eavesdropper is equivalent to an inertial eavesdropper of $\alpha_{I}=\frac{N_0}{N_0'}$. Both observers have the same capacity with Alice and Bob. 

We can use the security conditions for inertial observers to deduce the twin velocity which gives a secure channel for any noise asymmetry. We can even include the effect of motion or a gravitational Doppler shift in addition to a smaller noise. For an eavesdropper with a maximum Doppler factor $\alpha_{I}$ and a local noise $N_0'$, we just need to consider an inertial observer with $\alpha_{I}'=\alpha_I\frac{N_0}{N_0'}$. We will study the twin channel, but the same substitution can be done for the redshift channel. 

\begin{figure}[ht!]
\centering
\includegraphics{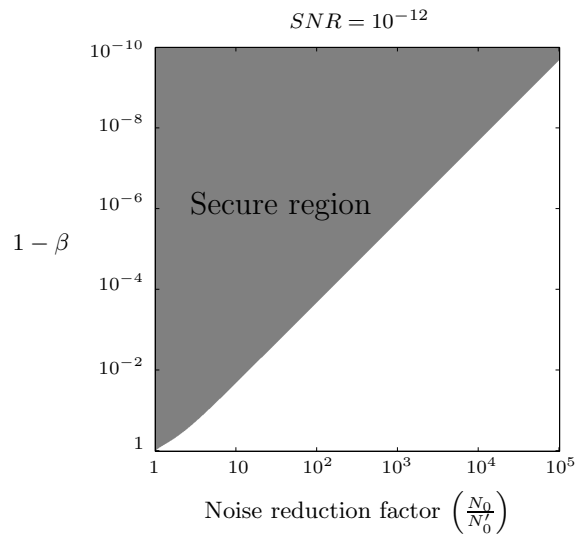}
\caption{\emph{Security conditions for low noise eavesdroppers}: Even if the eavesdropper has a noise $N_0'$ below the local noise $N_0$ of the twins, the twin channel is still secure for certain velocities. For a noise reduction factor $\frac{N_0}{N_0'}$, the graph shows all the velocities which permit a private channel. For high noise differences, the velocity $\beta$ is close to 1.}
\label{NoiseSecure}
\end{figure}

Figure \ref{NoiseSecure} shows the secure regions for an example with a weak signal. If we can bound the noise reduction factor $\frac{N_0}{N_0'}$ of the eavesdropper, we can deduce the velocity $\beta$ which gives us an advantage in the average capacity. We have not considered the case where $N_0'>N_0$, which will only improve security.

We can also give an approximate bound. We will consider the weak signal limit and suppose that $\alpha_{I}=\frac{N_0}{N_0'}$ is big so that $\beta$ must be close to 1 and $\alpha_{+}\gg 1$ and $\alpha_{-}\ll 1$. In that case, from (\ref{dopsqrt}), we can see $\beta=\frac{1-\alpha_{-}^2}{1+\alpha_{-}^2}\approx (1-\alpha_{-}^2)(1-\alpha_{-}^2)\approx 1-2\alpha_{-}^2=1-\frac{2}{\alpha_{+}^2}$. For these conditions, we know from (\ref{condsecweak}) that we have a secure channel if $\alpha_{+}>\alpha_{I}$. For a noise reduction factor $\frac{N_0}{N_0'}$ we can have a secure channel for velocities
\begin{equation}
\beta\ge 1-2\left(\frac{N_0'}{N_0}\right)^2.
\end{equation}

\newcommand{\noopsort}[1]{} \newcommand{\printfirst}[2]{#1}
  \newcommand{\singleletter}[1]{#1} \newcommand{\switchargs}[2]{#2#1}

\end{document}